\documentclass[final, aps, reprint, preprintnumbers, footinbib, showpacs, citeautoscript, superscriptaddress, nobalancelastpage, flushbottom]{revtex4-1}

\pdfoutput=1

\preprint{In Preparation}

\usepackage[T1]{fontenc}
\usepackage[english]{babel}

\usepackage{amsmath}
\usepackage{amssymb}

\usepackage{textcomp}
\usepackage{times}
\usepackage[notext]{stix}

\usepackage{braket}
\usepackage{graphicx}
\usepackage[dvipsnames]{xcolor}

\usepackage{subfigure}

\usepackage[colorlinks, urlcolor=blue, citecolor=blue, linkcolor=blue, pdfstartview=FitH]{hyperref}
\usepackage{hypcap}

\newcommand{\um}{\textmu{}m}
\newcommand{\uW}{\textmu{}W}

\begin{document}

\def\affiSOLAB{Spin Optics Laboratory, Saint~Petersburg State University, 198504 St.~Petersburg, Russia}
\def\affiTUDO{Experimentelle Physik 2, Technische Universit\"at Dortmund, D-44221 Dortmund, Germany}
\def\affiIOFFE{Ioffe Institute, Russian Academy of Sciences, 194021 St.~Petersburg, Russia}

\title{Increased sensitivity of spin noise spectroscopy using homodyne detection in \emph{n}-doped GaAs}

\author{M.~Yu.\ Petrov}
\affiliation{\affiSOLAB}
\author{A.~N.\ Kamenskii}
\affiliation{\affiTUDO}
\author{V.~S.\ Zapasskii}
\affiliation{\affiSOLAB}
\author{M.\ Bayer}
\affiliation{\affiTUDO}
\affiliation{\affiIOFFE}
\author{A.\ Greilich}
\affiliation{\affiTUDO}

\date{\today}

\begin{abstract}
We implement the homodyne detection scheme for an increase of the polarimetric sensitivity in spin noise spectroscopy.
Controlling the laser intensity of the local oscillator, which is guided around the sample and does not perturb the measured spin system, we are able to improve the signal-to-noise ratio. 
The opportunity of additional amplification of the measured signal strength allows us to reduce the probe laser intensity incident onto the sample and therefore to approach the non-perturbative regime.
The efficiency of this scheme with signal enhancement by more than a factor of 3 at low probe powers is demonstrated on bulk $n$-doped GaAs where the reduced electron-spin relaxation rate is shown experimentally.
Additionally, the control of the optical phase provides us with the possibility to switch between the measurement of Faraday rotation and ellipticity without changes in the optical setup.
\end{abstract}


\maketitle

\section{Introduction}

The most convincing demonstration of coherence in optics is known to be the classical effect of light interference.
Manifestations of this phenomenon in temporal domain underlie the effects of homodyning and heterodyning widely used nowadays as methods of detecting weak optical signals~\cite{KingstonBook}. 
Being phase sensitive, these methods allow one to get access to tiny variations of the light beam polarization~\cite{LevensonEesley79} and, in addition, may provide valuable information related to quantum properties of the light. 
Application of methods of quantum optics~\cite{SZBook, AndrewsBook, MetrologyBook} for detecting the spin state of charge carriers provides new possibilities for understanding spin-photon interfaces.

The standard method of analysis of spin dynamics is based on the pump and probe technique, where spin polarization is excited with a circularly polarized pump pulse and then measured with a linearly polarized probe that undergoes a rotation of its polarization plane due to the magneto-optical Faraday or Kerr effects~\cite{AwschBook}. 
Despite the linear polarization and the small excitation power, the probe beam still causes excitation of the system. 
A less perturbative measurement is performed when one probes the electron spin resonance in the transparency region of a semiconductor by measuring the fluctuations of the Faraday rotation at the frequency of the paramagnetic resonance~\cite{AZJETP81}. 
Mapping the spontaneous spin fluctuations in thermodynamic equilibrium onto the rotation fluctuations of the light polarization-plane is used in spin noise spectroscopy~\cite{CrookerNat04, OestreichPRL05, ZapasskiiPRL13, ZapasskiiAOP13, HuebnerPSSB14, SinitsynRPP16}. 
The mapping is governed by a spin-flip scattering~\cite{GPOS83} so that the transmitted light acquires Raman-shifted sidebands, such that the time-averaged intensity of the light field contains a contribution caused by an interference of the transmitted and scattered waves~\cite{GZOE15}.
As a result, the rotation angles to be measured hereby are very small and, thus, the polarimetric sensitivity should be as high as possible.

In general, the angle of Faraday rotation is proportional to the optical path length, which can be increased by placing the active medium into a cavity, either a macroscopic one~\cite{ZapPrzh11} or a microcavity~\cite{Kozlov13, PoltavtsevPRBR14, PoltavtsevPRB14}. 
Another method to increase the spectroscopic sensitivity is to increase the probe intensity sent through the sample while keeping the photon flux incident on the photodetector at a low level by diminishing the light intensity using a high polarization extinction (HPE) geometry~\cite{GlasenappPRB13}. 
The signal-to-noise ratio can be increased in HPE by orders of magnitude at the cost of a higher perturbation of the spin system by the increased power of the probe beam. 
In both cases, using either resonant cavity or high probe power, the interaction of light and matter is strongly increased, resulting in a stronger perturbation of the system. 
In practice, however, one wishes to reduce the probe power, keeping the spin system almost unperturbed.
In particular, weakly perturbative measurements might be used to probe the spin system in a cold atomic gas~\cite{MihailaPRA06}, in an electron gas at sub-Kelvin temperatures~\cite{FokinaPRB10} and in a charged cavity quantum-electrodynamics device~\cite{SmirnovPRB17}.

It should be noted, that the Faraday rotation can also be enhanced by coupling a carrier spin to a magnetic particle, like in a diluted magnetic semiconductor~\cite{CronenbergerNC15}. 
Furthermore, as one tests the spin system using a noisy light field, a certain benefit could be achieved by probing with non-classical light, having a reduced level of photon shot noise, requiring, however, an elaborated laser setup to control the light properties~\cite{LuciveroPRA16}.

The homodyne measurement of the Faraday rotation, used in this paper, has been applied to combine benefits of the geometry of HPE detection, while simultaneously keeping the probe power as low as possible. 
Demonstrated previously in an improved pump-probe version~\cite{LaForgeAPL07, LaForgeRSI08}, a similar method was adopted recently to spin noise spectroscopy of $n$-doped GaAs placed in a microcavity in Ref.~\onlinecite{CronenbergerRSI16}, with a focus on the realization of a quantum-limited homodyne and heterodyne detection in order to extend the detection-frequency range. 
The general idea of such a measurement, shown particularly by Cronenberger and Scalbert (Ref.~\onlinecite{CronenbergerRSI16}), is based on the use of an interferometric setup, where phase fluctuations encoded in a weak spin noise signal are measured by mixing the scattered light with a strong reference beam, the so-called local oscillator (LO), which is not interacting with the spin system. 
The increase of the signal-to-noise ratio, in this case, is governed by the fact that the LO might be chosen as strong as required to overcome the stray noises of electronics, thus limiting the efficiency of measurements only by the level of the photon shot noise and the dynamic range of the photodetector.

Here, we utilize a balanced homodyne technique using a Mach-Zehnder interferometer to measure the spin noise of electrons at the edge of the Fermi sea in a $n$-doped GaAs epilayer.  
We implement an optical path length stabilization so that this method provides benefits to map the spin fluctuations in different quadratures and allowed to perform long-time accumulation of the spin noise signal.
The electron-spin relaxation rates are thus measured at levels of perturbation varied by three orders of magnitude in excitation density.

\section{Results}
\subsection{Conventional spin noise experiment}
\begin{figure}[t]
\subfigure{\includegraphics[clip]{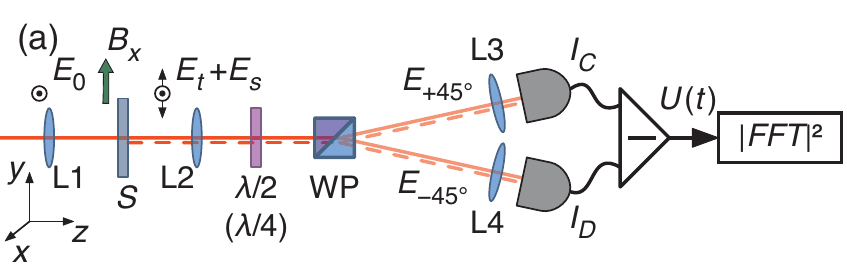}\label{fig:One:a}}
\hbox{\subfigure{\includegraphics[clip]{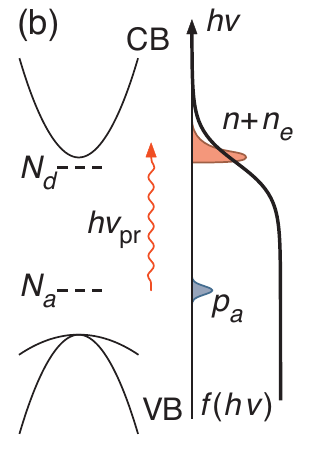}\label{fig:One:b}}%
\subfigure{\includegraphics[clip,width=54mm]{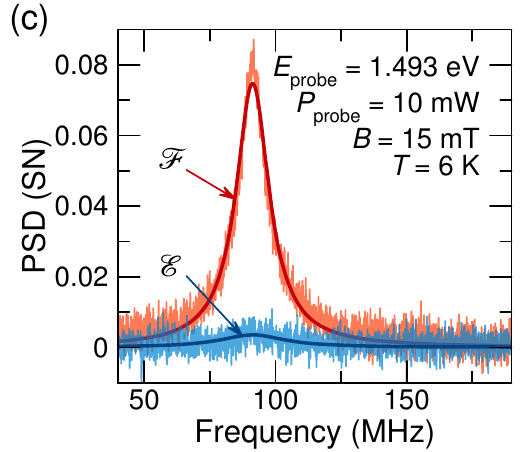}\label{fig:One:c}}}
\caption{Measuring the spin noise using Faraday rotation and ellipticity.
(a) Schematic of the experimental setup: the sample is probed with continuous wave excitation whose wavelength is tuned below the band-to-band absorption into the band gap of GaAs ($\lambda_\mathrm{pr} = 830$~nm, $E_\mathrm{probe} = 1.493$~eV). A classical polarimeter consisting of a half-wave plate ($\lambda/2$) for the Faraday rotation measurement or a quarter-wave plate ($\lambda/4$) for the ellipticity measurement, followed by a Wollaston prism (WP) and a balanced photoreceiver. The difference photocurrent ($I^- = I_C - I_D$) generated there is converted into a voltage $U(t)$ the frequency response of which is obtained by measuring its power spectral density.
(b) Energy scheme of $n$-type GaAs and carrier distribution function with carrier concentrations as the function of energy. 
(c) Power spectral density of the electron spin noise detected by Faraday rotation ($\mathscr{F}$) and ellipticity ($\mathscr{E}$), measured in a magnetic field $B_x = 15$~mT at a sample temperature $T = 6$~K. The probe power at the sample is $10$~mW. The data (noisy curves) and their Lorentzian fits (solid lines) are shown in units of the photon shot noise.}\label{fig:One}
\end{figure}

We start our analysis from a description of the conventional scheme used in spin noise spectroscopy, schematically shown in Fig.~\ref{fig:One:a}.
The sample (S) is exposed to coherent monochromatic light emitted by a single-frequency Ti:Sapphire laser propagating along the $z$~direction and polarized linearly along the $x$~axis. 
A single spatial mode of the beam, further referred to as $E_0$, is selected with a single-mode, polarization maintaining fiber. 
The light is focused on the sample and collimated with $F=100$~mm achromatic doublets selected to optimize the focal depth according to the sample thickness [L1 and L2 in Fig.~\ref{fig:One:a}]. 
When passing through the medium, the light undergoes scattering on the fluctuations of the spin density leading to the appearance of secondary waves: $E_t$ (transmitted beam) of the same linear polarization as $E_0$ and the additional mode, $E_s$, (scattered beam) having the orthogonal linear polarization.
Note that a phase shift between the polarization components of the electromagnetic wave in the modes $E_t$ and $E_s$ may occur due to a difference in absorption of the circular left and right polarizations in the sample.

A standard polarimetric setup is based on analysis of the difference of the photosignals in the two arms behind a polarizing beamsplitter or a Wollaston prism (WP). 
The polarization before the WP is oriented at $45$\textdegree{} relative to the $x$~axis, to provide the equal intensities of the transmitted beams, which is achieved by an appropriate rotation of the half-wave plate in front of the WP. 
When probing the ellipticity, the half-wave retarder is replaced by a quarter-wave plate. 
The excess intensity noise is suppressed by balanced optical bridge detection where the photodiodes are wired in series to produce a differential current that is converted to a voltage and amplified in a relatively wide spectral band from 0.1 to 650~MHz~\cite{CommentBHD}. 
The voltage fluctuations, amplified with two low-noise voltage pre-amplifiers both provided 20~dB voltage gain, are then digitized and Fourier-transformed in 1~GHz frequency band by using the real-time accumulation of the noise power spectral density (PSD), see Ref.~\onlinecite{CrookerPRL10} for more technical details.

We study a bulk layer of negatively doped GaAs, where the electron concentration is close to the metal-insulator transition~\cite{CrookerPRB09}.
At low temperature, a fraction of the donor electrons is thermally excited into the conduction band, providing an electron density $N_e \simeq 3.7\times10^{16}$~cm$^{-3}$ at $T = 10$~K. 
When tuning the probe into the band gap, residual absorption of light may occur due to the carbon acceptor band~\cite{BrozelChap} as shown in Fig.~\ref{fig:One:b}.

The spin noise spectra are measured by switching the magnetic fields between $B_x = 15$~mT [with corresponding frequency spectrum $\mathcal{P}_1(\nu)$] and $B_x = 100$~mT [$\mathcal{P}_2(\nu)$], where $\nu$ is the frequency. 
The PSD of the spin noise is shown in units of the shot noise:
\begin{equation}
\mathrm{PSD\;(SN)} = \frac{\mathcal{P}_1(\nu) - \mathcal{P}_2(\nu)}{\mathcal{P}_2(\nu) + \mathcal{P}_e(\nu)}.
\label{eq_PSDSN}
\end{equation}
Here, $\mathcal{P}_e(\nu)$ is the electronic noise of the photoreceiver amplifier and the recording instruments.

\begin{figure*}[t]
\subfigure{\includegraphics[width=\columnwidth]{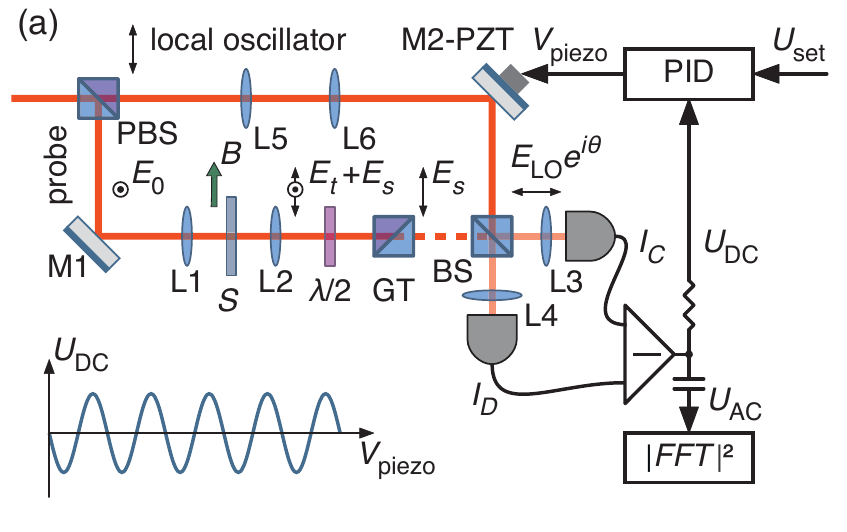}\label{fig:Two:a}}
\subfigure{\includegraphics[width=\columnwidth]{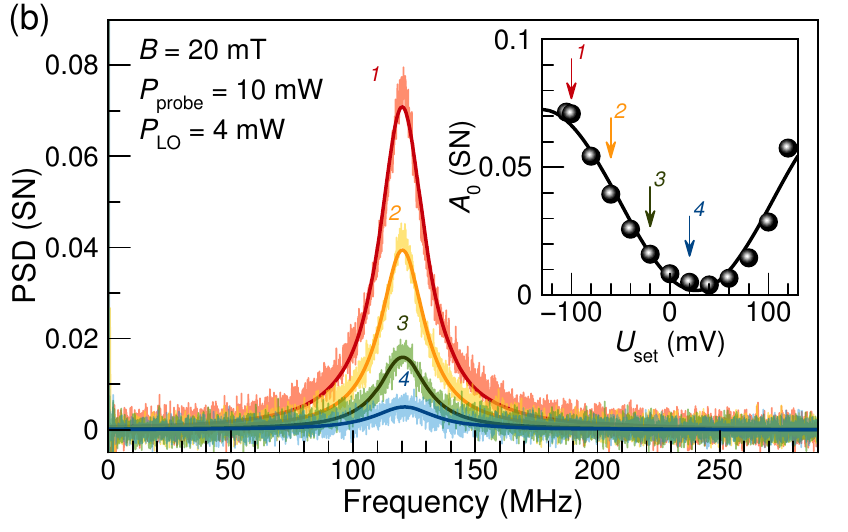}\label{fig:Two:b}}
\caption{Measuring the spin noise using a balanced homodyne technique.
(a) Schematic of the homodyne detection setup. The single-mode laser beam is split at the input of the Mach-Zehnder interferometer into the two interferometer arms by a polarizing beam splitter. The sample is placed in the probe arm where the spin noise is monitored in the transmission geometry by illuminating it with the probe beam in mode $E_0$ that is linearly polarized along the $x$~axis. The transmitted and scattered light have orthogonal linear polarizations corresponding to the modes $E_t$ and $E_s$, respectively. The half-wave plate ($\lambda/2$) and the Glan-Taylor polarizer (GT) are used to filter out the electric-field mode $E_t$. The passed scattered light mode $E_s$ and the mode of light in the reference arm of the interferometer (local oscillator) are sent to the input of the 50:50 non-polarizing beam splitter (BS). The interference of the electric field mode $E_s$ and the local oscillator $E_\mathrm{LO}$ results in the photocurrents $I_C$ and $I_D$ in the balanced photoreceiver, where $I^- = I_C - I_D$ is converted into the voltage signal $U(t)$. Two components of $U(t)$ are analyzed: the AC component is analyzed using a real-time Fourier transformed acquisition similar to the traditional method, and the DC component is sent to the error input of the PID control loop used to adjust the voltage $V_\mathrm{piezo}$. Thereby the relative optical phase shift between the two arms of the interferometer $\theta$ is maintained by tuning the piezo-actuated mirror (M2-PZT) to the set point $U_\mathrm{set}$. The inset shows a schematic of the $U_\mathrm{DC}$ versus $V_\mathrm{piezo}$ dependence.
(b) Spin noise spectra (noisy curves) and their Lorentzian fits (solid lines) measured in a linear combination of the $E_s$ field quadratures by varying the phase $\theta$. A continuous evolution of the spin noise of Faraday rotation to the noise of ellipticity is obtained (curves 1--4). The inset shows the amplitude of the spin noise peak extracted from fitting. The arrows indicate the points where the curves presented in the panel are measured.}
\label{fig:Two}
\end{figure*}

Figure~\ref{fig:One:c} demonstrates a typical spin noise signal measured in Faraday rotation and ellipticity configurations. 
The presence of the ellipticity component is related to the residual absorption, which leads to increased perturbation of the electron spin system and accelerated relaxation dynamics. 
Such measurement, therefore, cannot be considered as completely non-perturbative and requires a very small probe power.

The results of the measurements presented in Fig.~\ref{fig:One:c} can be treated analytically. 
The secondary wave $E_s$ appears due to the scattering of the probe light on the fluctuations of the spin density.
On long-time average, its amplitude is zero. However, a rigorous calculation using the `beam splitter' model (see Appendix~\ref{sec:Methods}) shows that the signal may be detected as the interference of the directly transmitted beam $E_t$ and $E_s$. 
The difference signal at the output of the photodetector is given by:
\begin{equation}
I_\mathscr{F}^- = \eta (E_t^*E_s + E_s^*E_t)	
\label{eq_Curr_Faraday}
\end{equation}
for Faraday rotation, and for ellipticity it is given by:
\begin{equation}
I_\mathscr{E}^- = i\eta (E_t^*E_s - E_s^*E_t),
\label{eq_Curr_Ellipticity}
\end{equation}
where $\eta$ accounts for the spectral sensitivity of the photo-detection.

Note, that the spin noise is encrypted in $E_s$ only, while the electric field of the transmitted beam can only be influenced by fluctuations of the charges~\cite{YugovaPRB09}. 
Because of that, charge fluctuations are often detected in the spin noise of charge-tunable quantum dots~\cite{KuhlmanNP13}. 
Therefore, the transmitted beam can be replaced by any light field provided its coherence (i.e., the optical phase synchronization) is maintained with the probe during the times of spin noise signal accumulation, as considered in the following.

\subsection{Spin noise in homodyne detection}

In the limit of $E_t \gg E_s$, Eqs.~\eqref{eq_Curr_Faraday} and \eqref{eq_Curr_Ellipticity} provide information on the noise of the real and imaginary Hermitian quadratures of the scattered field that can be measured with a phase-sensitive detection scheme, such as balanced homodyne. 
To implement this method, we perform measurements using the setup shown in Fig.~\ref{fig:Two:a}. 
The sample is placed in one arm of the Mach-Zehnder interferometer, where the polarimetric analysis of the scattered light is done using the HPE geometry. 
In this case, the half-wave plate is placed in front of the Glan-Taylor (GT) polarizer providing an extinction ratio of 1:10000. 
The homodyne scheme uses a non-polarizing 50:50 beam splitter (BS) to combine a negligibly small-in-amplitude field $E_s$ with the coherent field of the LO. 
Importantly, the LO is purified with an additional GT, and the wavefront of the LO mode is carefully collimated with a pair of achromatic doublets [L5 and L6 in Fig.~\ref{fig:Two:a}]. 
To provide a good spatial overlap of the modes, the light mode $E_t$ is used to obtain an interference pattern and, after the initial arrangements, the transmitted light is filtered out by rotating the half-wave plate [Fig.~\ref{fig:Two:a}]. 
The inputs of the optical bridge are connected to the output ports of the BS.
As in the conventional detection scheme, the spin noise is monitored in the frequency domain by the accumulation of the power spectrum.

To obtain information on the field quadratures of $E_s$, the phase of the LO is constantly controlled during signal accumulation.
This is done by adjusting the optical path length in the LO arm with the piezo-actuated mirror holder (M2-PZT).
In order to implement the proportional-integral-differential (PID) control of the phase stabilization, we slightly detune the half-wave plate to allow a tiny part of $E_t$ to be transmitted into the detection channels.
An additional DC current monitor of the photoreceiver is used to detect the low-frequency difference signal.
The control voltage sent to the piezo actuator is proportional to the error signal detected as difference between the low-frequency balanced output and a set point, see Fig.~\ref{fig:Two:a}. 
In addition, implementing the phase stabilization scheme allows one to reduce the low-frequency noise (below 50~MHz) in the spin noise power spectrum [see Fig.~\ref{fig:Two:b}], compared with a one-port homodyne measurement~\cite{CronenbergerRSI16}, because the balanced homodyne detection can be made insensitive to LO quadrature-phase noise~\cite{YuenOL83, SchumakerOL84}.

Figure~\ref{fig:Two:b} represents the power spectra of the spin noise measured by the homodyne technique at various phase set points, i.e., at different relative optical phases of the LO with respect to $E_t$, and, correspondingly, to $E_0$ and $E_s$. 
As one can see from the figure, the spin noise power drastically depends on the phase, see inset in Fig.~\ref{fig:Two:b}.
To understand the observed behavior, we use a model in which all important optical elements are taken into account, see Appendix~\ref{sec:Methods}.
Since the LO-field is a strong coherent field, its state can be expressed as $E_\mathrm{LO}e^{i\theta}$.
Accounting for the phase tuning, a variable optical phase shift $\theta$ can be added to without any restriction so that the inputs of BS are coupled to the analyzing field $E_s$ and the LO field. 
The difference current of the photodetector output can be written as:
\begin{equation}
I^-_\mathrm{HD} = K \eta E_\mathrm{LO} (E_s^* e^{i\phi} + E_s e^{-i\phi}),	
\label{eq_Curr_Homodyne}
\end{equation}
where $K$ characterizes the spatial overlap of the LO and signal modes and $\phi = \theta + \pi/2$.
Readily, at $\phi = 0$ and $\phi = \pi/2$ the Faraday rotation and ellipticity noise are measured, respectively. 
These phase points correspond to the curves 1 and 4 in the inset to the Fig.~\ref{fig:Two:b} and reproduce quite well the amplitude of both curves in Fig.~\ref{fig:One:c} when measured under the same excitation conditions (10~mW at the sample) and laser power at the photodiodes (2~mW per each photodiode).

There are several advantages of the homodyne detection scheme over conventional polarimetry. 
First, the ability to make the LO intensity $P_\mathrm{LO}$ as strong as possible while decreaing the power of the probe light may reduce the impact of nonlinear processes due to the absorption of the probe beam. 
The only limitation is the dynamic range of the photodiodes, as we will show further.
Second, the homodyne detection is able to measure not only the forward-scattered light but also other harmonics that might propagate outside of the mode of the transmitted beam~\cite{GPOS83}.
Third, there is the ability to measure arbitrary linear combinations of the electromagnetic-field quadratures in the phase space.

\begin{figure*}
\subfigure{\includegraphics[width=\columnwidth]{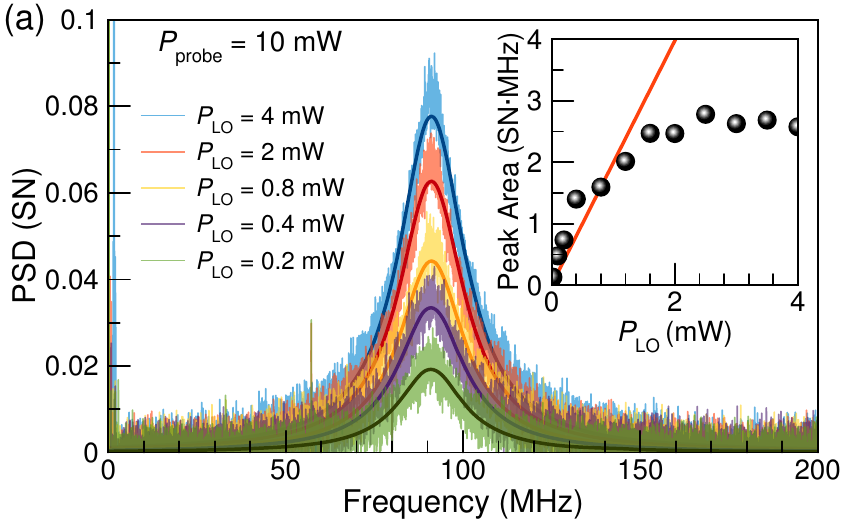}\label{fig:Three:a}}\
\subfigure{\includegraphics[width=\columnwidth]{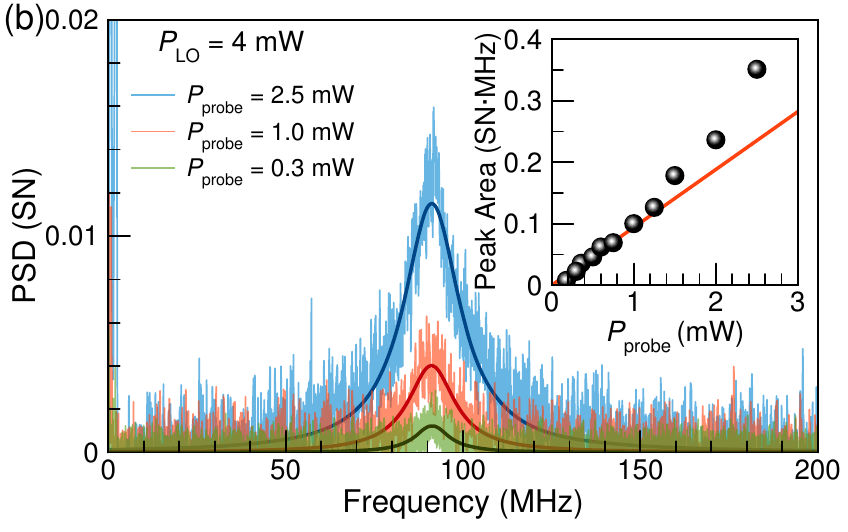}\label{fig:Three:b}}\\
\subfigure{\includegraphics[width=\columnwidth]{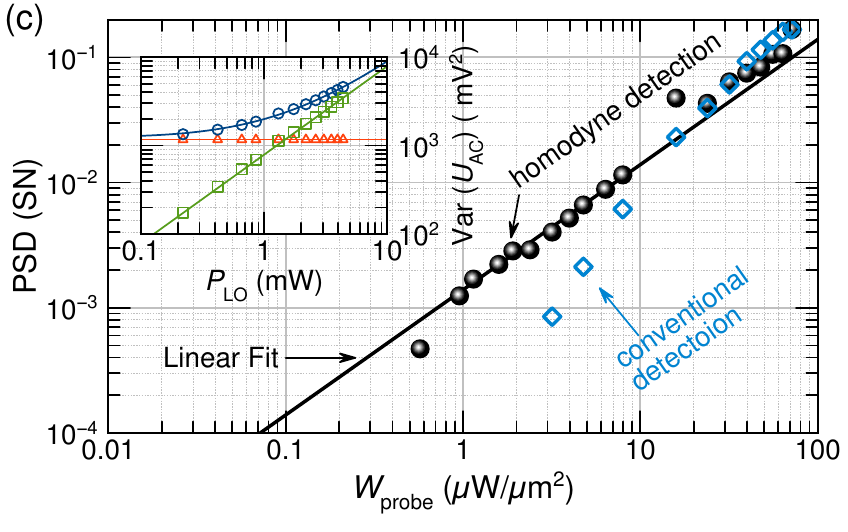}\label{fig:Three:c}}\
\subfigure{\includegraphics[width=\columnwidth]{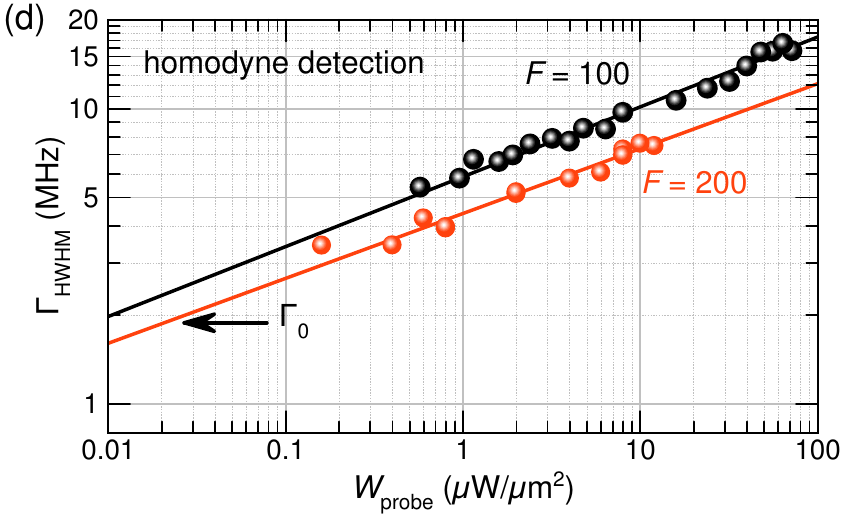}\label{fig:Three:d}}
\caption{Analysis of the perturbation of the spin system using homodyne detection of spin noise.
(a) Power spectral density of the Faraday rotation noise measured at fixed probe power using the homodyne technique at varying powers of the local oscillator. The inset shows the power dependence of the integrated spin noise power (symbols), a linear fit to its area is shown by the red line.
(b) Spin noise spectra measured at varying probe power and constant power of the local oscillator. The inset displays the area of the spin noise power (symbols) extracted from fitting. A linear fit of the curve is provided by the red line.
(c) Amplitudes of the spin noise PSD peaks versus probe power density measured in the conventional detection scheme (blue squares) and with the homodyne detection (black circles). A linear fit is given by the black line. The inset shows the variance of the output voltage of the photoreceiver used for balanced homodyne detection, including total noise (blue circles), electronic noise (red triangles), and extracted shot noise (green squares).
(d) Electron-spin relaxation rate versus probe excitation power density, extracted from fitting the PSD curves measured with $F=100$~mm (black circles) and $F=200$~mm (red circles) lenses in the probing arm. Lines are fits using a power function: $\Gamma_\mathrm{HWHM} \propto W_\mathrm{probe}^{0.22}$. The value of the spin relaxation rate in darkness ($\Gamma_0$) is indicated.}
\label{fig:Three}
\end{figure*}

\subsection{Analysis of spin system perturbation}

The main difference between the conventional and the homodyne detection schemes is that in the latter one the spin noise is normalized to the shot noise of the LO field, which means, in an ideal situation, that the signal-to-noise ratio can be increased infinitely. 
However, this is not the case in reality because of several reasons accounting for the nonlinearities.
First, let us examine the nonlinearity of the detection scheme. In Fig.~\ref{fig:Three:a} the spin noise is measured at a constant power of the incident probe and varying LO intensity. 
As seen, the power spectra of the spin noise might be effectively increased by varying the LO intensity. 
However, at small intensities only a linear behavior according to Eq.~\eqref{eq_Curr_Homodyne} is obtained. 
Saturation of the amplifier of the photoreceiver occurs around 2\,mW, meaning that dependence $\eta(E_\mathrm{LO})$ should not be neglected. 
One can consider the measured LO power dependence of the spin noise power spectrum as a merit of the photoreceiver nonlinearity.

Second, the nonlinearity of the measured spin system reveals itself in the deviation of the PSD area from a linear probe power dependence, see Fig.~\ref{fig:Three:b} and contained inset. 
This means that the spin system becomes more ``noisy'' by an additional carrier excitation caused by the probe itself. 
The analysis also reveals a broadening of the spectra, $\Gamma_\mathrm{HWHM}$, obtained as half-width at half-maximum (HWHM) of the spin noise peak, with probe power, as shown in  Fig.~\ref{fig:Three:d}, corresponding to an increase of the spin relaxation rate. 
The laser power on the photodiodes, in this case, is solely given by the power of LO, which is kept constant at 4~mW. 
The transmitted probe power is completely blocked by the Glan-Taylor polarizer after the sample, and the power of the scattered light component is negligible in comparison to the power of the LO.

In order to analyze the perturbation of the spin system as a function of probe power, we performed a series of comparative measurements using the conventional and the homodyne detection schemes. 
The amplitudes of the spin noise measured by both methods at the same excitation
densities of the probe are compared in Fig.~\ref{fig:Three:c}. 
As seen in the log-log scale presentation, in the range from 10 to 100~\uW/\um$^2$ ($P_\mathrm{probe}>1$~mW) both methods deliver the same PSD signal. 
However, in the range from 1 to 10~\uW/\um$^2$ the conventional detection scheme could not be used anymore because the electronic noise $P_e(\nu)$ becomes dominant over the photon shot noise in our diodes. 
As a consequence, no spin noise signal can be recorded, independent of the accumulation time.
In this case, the homodyne detection provides a good solution, as the constant power of the LO at the diodes keeps them at a level, where the photon shot noise is always above the electronic noise. 
This makes it possible to accumulate very weak noise signals and leads to better agreement with the theoretically expected linear dependence given by Eq.~\eqref{eq_Curr_Homodyne}. 
At densities of excitation below 1~\uW/\um$^2$, the PSD accumulation requires more than an hour, which we decided to be the limit of the accumulation time.

To evaluate the power densities required for efficient conventional detection and homodyne detection of spin noise, we plot the characteristic function of our photoreceiver in the inset of Fig.~\ref{fig:Three:c}.
The data are measured in the absence (for electronic noise) and in presence of constant illumination of the LO light (photon + electronic noise), and further integrated over the broadband frequency range to compute the net power.
Here, one can see that the photon shot noise becomes dominant in relatively narrow range of light powers from $P_\mathrm{LO} = 2$ to $P_\mathrm{LO} = 4$~mW that limits the dynamic range of the homodyne detection.
Note, the data on the Fig.~\ref{fig:Three:c} are normalized to the total noise (photon + electronic). 
If the data are normalized only on the photon shot noise, the conventional scheme shows similar linear dependence as the homodyne one but the efficient measurement could be performed only at relatively large probe power.

It is important to evaluate also the spin relaxation rates for the condition of low optical excitation. 
To that end, we performed a series of measurements with lenses of different focal length, used to focus the probe beam onto the sample. 
The spot sizes are measured by a beam profiler, from which beam diameters of $20$~\um{} and $40$~\um{} are obtained for lenses with focal lengths of $F = 100$~mm and $F = 200$~mm, correspondingly. 
By increasing the focal length twice we reduced the optical density by an additional order of
magnitude. 
In addition, the bigger spot size reduces the effect of transit-time broadening of the spectrum, which was estimated to be of the order of 10~\um{} in our sample, see Ref.~\cite{CrookerPRB09}. 
Figure~\ref{fig:Three:d} summarizes the two series of experiments. 
As seen there, $\Gamma_\mathrm{HWHM}$ drops with decreasing probe power following a power function with exponent $0.22$. 
A spin relaxation time of $\tau_s = (2\pi \Gamma_\mathrm{HWHM})^{-1}= 52$~ns is obtained at the lowest density of excitation, which is comparable with measurements of the same sample using the Hanle effect~\cite{CrookerPRB09}. 
In contrast, in an extended pump-probe experiment~\cite{BelykhPRB16, BelykhPRB17}, the spin relaxation time reaches values of $(2\pi\Gamma_0)^{-1}=90$~ns under similar optical excitation conditions of the same sample. 
Note that in this case, the optically excited electron spins evolve in the darkness between the arrival of the pump and probe pulses. 
Fitting the dependencies presented in Fig.~\ref{fig:Three:d} shows that in this case the value of $\Gamma_\mathrm{HWHM}$ extrapolated to $\Gamma_0$ corresponds to probe powers about one order of magnitude lower than in our experiments.
\looseness=-1
\raggedbottom

\section{Discussion}

The possibility to control the optical phase in spin noise measurements performed with the homodyne detection is beneficial when the spin system is unavoidably perturbed by the measurement.
By changing only the path length for the reference light beam that does not interact with the spin system, the full phase-space image of the spin noise can be reconstructed without any replacements of the optical elements.
This is especially important for systems in which we are forced to detect the spin noise perturbatively by resonant excitation, e.g. in ensembles of quantum dots, where the spin system is probed inside the inhomogeneously broadened absorption line.
In particular, the Faraday rotation is zero in the precisely resonant excitation conditions while the fluctuations of spins subject to the tail of other spectral lines in the ensemble are probed.
On the contrary, the noise of ellipticity reveals optical properties of spectral lines, which resonant frequencies are spectrally spaced within the homogeneous broadening range. 
These are the specific features explored by optical spectroscopy of spin noise (see Ref.~\onlinecite{ZapasskiiPRL13} for details).

An important step in that respect is the implementation of a phase stabilization loop, which allowed us to remove the excess noises at frequencies below $50$~MHz. 
The improved stability made it feasible to easily accumulate spin noise during long time periods and thus to accumulate the signal when the probe  perturbs the spin system only slightly, thus approaching the regimes of measurement of the intrinsic spin relaxation time. 
In the studied system of $n$-doped bulk GaAs, this time is, however, still limited by the absorption tail so that the intrinsic time might be longer than the one measured with spin noise.
\looseness=-1

One should also note that there are optimal conditions of probing spin noise for an ensemble of spins.
Lucivero \emph{et~al.}~(Ref.~\onlinecite{LuciveroPRA17}) performed experimental measurements and statistical analysis to evaluate the global standard quantum limits defining the limiting sensitivity of spin noise spectroscopy.
They have shown that in optically probed hot atomic vapors of $^{85}$Rb the limiting sensitivity could be achieved at atom density of about $7\times10^{12}$~cm$^{-3}$ and probe power of about 7~mW.
In these conditions, the homodyne detection could not help and the polarization-squeezed probe beam surpasses the global standard quantum limit for this system.
In case of the electron gas, like in $n$-GaAs studied in this work, the functional dependence of the power broadening of the spin noise spectrum is more complicated than in an atomic system.
Therefore, quantitative validation of optimal conditions could be performed in the further works with the help of conventional detection, including HPE at high-power conditions and using the homodyne detection at low-power conditions.

As already mentioned, further improvements in measurement sensitivity may be achieved by analyzing the additional scattering modes that appear in a system where the spins are spatially localized, like in quantum dots, for which the effective size of the scatterers is smaller than the wavelength of the probe light~\cite{GPOS83}.
In this case, mode shaping of the LO could potentially help to probe the signal modes that do not propagate along the transmitted or reflected rays~\cite{CronenbergerRSI16}. 
In our case, however, we do not see any spin noise signal outside of the aperture of the forward scattered light due to the large probed volume comparable to the wavelength of light.
At the same time, the use of signal modes propagating outside the aperture of the reference beam can be hampered by kinetic motion of spins possible at this level of doping~\cite{KozlovPRA18}.

\section{Conclusion}

To summarize, we implemented polarization sensitive interferometry for the detection of Faraday rotation and ellipticity.
The spin noise of $n$-doped GaAs was measured in different quadratures by variation of the path difference in the arms of the Mach-Zehnder interferometer. 
Regimes of perturbation of the spin system are analyzed by measuring the electron spin relaxation time at excitation powers varied over several orders of magnitude while probing in the transparency region of GaAs.
The quantitative analysis shows that for all reasonable intensities of the probe we find an amplification of the sensitivity of the homodyne detection scheme over the one of a conventional 45~degrees polarimetric setup.
The obtained signal to noise ratio is found to be always larger in the former scheme and is limited only by the finite dynamic range of the photoreceiver. 
These findings might be used to implement weak measurements of spin dynamics on the nanoscale.

\emph{Note Added.} After submittal, we became aware of a similar research done on isotopically enriched rubidium vapors where the homodyne detection of spin noise is done in the low-frequency range~\cite{Sterin18}.

\acknowledgements

The authors are grateful to D.~S.\ Smirnov and V.~V.\ Belykh for valuable discussion and thank S.~A.\ Crooker for providing the sample.
This research was done in the frame of the International Collaborative Research Center supported by the Deutsche Forschungsgemeinschaft TRR 160 (project A5) and the Russian Foundation for Basic Research (Project No.~15-52-12013).
MYP thanks the Resource Center ``Applied Aerodynamics'' of SPSU for technical support with design of the optical enclosure and acknowledges financial support of Saint-Petersburg State University (research grant 11.34.2.2012).

\appendix

\section{Description of conventional and balanced homodyne detection}
\label{sec:Methods}

Here, we provide a rigorous description of the optical field measurements described in the main text.
Suppose $\mathbf{E}_0$ be an electric-field of the probe light polarized along the $x$~axis and traveling along $z$.
Being transmitted through the sample the probe beam induces the polarization of the medium and exhibits scattering so that the transmitted light field has a secondary wave $\delta \mathbf{E} (t) \propto \mathbf{P}(t)$ proportional to the dielectric polarization which, in turn, has two contributions: $\mathbf{P}(t) = \mathbf{P}_x(t) + \mathbf{P}_y(t)$. 
The first term, $\mathbf{P}_x(t)$, is polarized as $\mathbf{E}_0$ and does not contain a contribution from spin fluctuations, the second term, $\mathbf{P}_y(t)$, proportional to the fluctuating magnetization has orthogonal linear polarization~\cite{YugovaPRB09}.
Both components are linear in the amplitude of the incident field and contribute to the resonance fluorescence and to the Raman scattering and spin noise, respectively~\cite{GZOE15}.
Therefore the electric field of the light incident on the polarization analyzer can be expressed as
\begin{equation}
	\mathbf{E}_\mathrm{transmit} = T E_0 \mathbf{e}_x + R E_0 \mathbf{e}_y
\end{equation}
where $\mathbf{e}_{x,y}$ are the unit vectors and $T$ and $R$ are the complex-value coefficients describing attenuation and light-matter interaction of the probe beam with the medium.
Note that $T\lesssim 1$ and $R \ll T$ in a real experiment, therefore $E_t \gg E_s$ where $E_t = T E_0$ and $E_s = R E_0$.

\begin{figure}[t]
\includegraphics[width=\columnwidth]{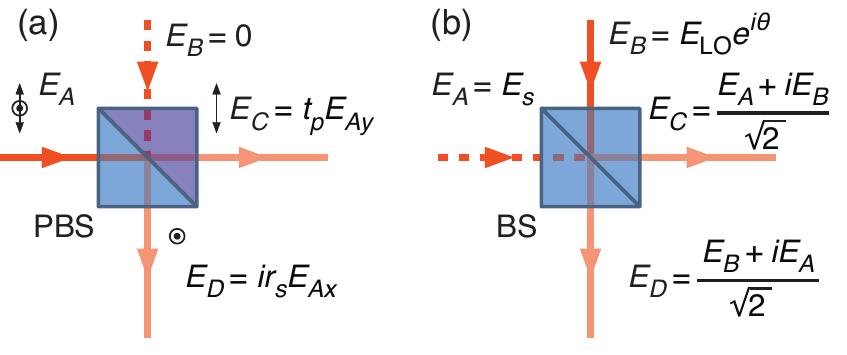}
\subfigure{\label{fig:Four:a}}
\subfigure{\label{fig:Four:b}}
\caption{Schematic diagrams of (a) a polarizing beam splitter and (b) a non-polarizing beam splitter.
}\label{fig:Four}
\end{figure}

In the standard detection scheme shown in Fig.~\ref{fig:One:a}, the polarization of the transmitted light is rotated by the angle of 45\textdegree\ using a half-wave plate and then split at the Wollaston polarizer for which we use a model of the polarizing beam splitter (PBS) where only one input port is coupled to the light [see Fig.~\ref{fig:Four:a}].
The PBS mixes the light modes input to ports $A$ and $B$ ($E_A$ and $E_B$) so that the output ports $C$ and $D$ ($E_C$ and $E_D$) are related to the inputs as:
\begin{equation}
	\begin{pmatrix}
		{E_C}_x\\ {E_C}_y\\ {E_D}_x\\ {E_D}_y
	\end{pmatrix}
	=
	\begin{pmatrix}
		t_p  & 0    & ir_p' & 0\\
		0    & t_s  & 0     & ir_s'\\
		ir_p & 0    & t_p'  & 0\\
		0    & ir_s & 0     & t_s'
	\end{pmatrix}
	\begin{pmatrix}
		{E_A}_x\\ {E_A}_y\\ {E_B}_x\\ {E_B}_y
	\end{pmatrix}
\end{equation}
where $t_{s,p}$, $r_{s,p}$, $t_{s,p}'$, and $r_{s,p}'$ are the transmission and reflection coefficients of the PBS for the light of $s$ and $p$ polarization that enter at the ports $A$ and $B$, respectively.
For the ideal lossless PBS: $t_p = t_p' = r_s = r_s' = 1$, $t_s = t_s' = r_p = r_p' = 0$, and the calcite-made WP gives a reasonably good approximation providing extinction ratio >10000:1 for both output beams.

Taking ${E_B}_{x,y} = 0$, the PBS outputs are related to $E_t$ and $E_s$:
\looseness=-1
\begin{equation}
	E_C = \frac{1}{\sqrt{2}}\bigl(E_t - E_s\bigr),\quad
	E_D = \frac{i}{\sqrt{2}}\bigl(E_t + E_s\bigr),
\end{equation}
where the reflected beam at port $D$ is phase shifted by $\pi/2$ relative to the transmitted beam at port $C$.

To measure the ellipticity, the half-wave retarder is replaced by a quarter-wave plate oriented such that the $\sigma^+$ and $\sigma^-$
components are translated into the basis of linear polarization and split at the PBS. In this case, the output modes of the PBS read as:
\begin{equation}
	E_C = \frac{(1-i)E_t + (1+i)E_s}{2},\quad
	E_D = \frac{(1+i)E_t + (1-i)E_s}{2}.
\end{equation}

The photo-currents generated at the two photodetectors are proportional to the average number of incoming photons in the corresponding channels, i.e., to the light intensity, therefore the difference current at the output of the balanced photoreceiver is given by $I^- = \eta \left(E_C^*E_C - E_D^*E_D\right)$, where $\eta$ is a constant describing the photon flux to voltage-drop conversion of the photodetectors. 
The difference signal is therefore expressed as the interference of transmitted and scattered light modes by:
\begin{equation}
I^-_\mathscr{F} = \eta (E_t^*E_s + E_s^*E_t),
\label{eq:methods:faraday_difference_current}
\end{equation}
for Faraday rotation, and for the ellipticity it reads as:
\begin{equation}
I^-_\mathscr{E} = i \eta (E_t^*E_s - E_s^*E_t).
\label{eq:methods:ellipticity_difference_current}
\end{equation}

In the homodyne geometry [Fig.~\ref{fig:Two:a}], the beam attenuator consisting of the half-wave plate and the Glan-Taylor polarizer is used to split the mode of scattered light, again providing an extinction ratio exceeding 10000:1. 
Then, the signal arrives at one input port of the 50:50 non-polarizing beam splitter (BS) and  the other input port is coupled to the LO [Fig.~\ref{fig:Four:b}].
Considering an ideal lossless BS where $t_{p,s} = t_{p,s}' = r_{p,s} = r_{p,s}' = 1/\sqrt{2}$, the beam splitter output modes are given by
\looseness=-1
\begin{equation}
	E_C = \frac{E_A + i E_B}{\sqrt{2}},\quad
	E_D = \frac{E_B + i E_A}{\sqrt{2}}.
\end{equation}
In this case, the difference signal at the output of the balanced photoreceiver can be expressed as:
\begin{equation}
I^-_\mathrm{HD} = i\eta (E_A^*E_B - E_B^*E_A).
\label{eq:methods:homodyne_difference_current}
\end{equation}

Let port $B$ be coupled to the field $E_B = E_\mathrm{LO} e^{i\theta}$ where $E_\mathrm{LO}$ and $\theta$ are the real-value amplitude and a variable phase shift.
Taking $E_A = E_s$, $\phi = \theta + \pi/2$, and $K<1$ as a numerical characteristic of the spatial overlap of modes $E_\mathrm{LO}$ and $E_s$, Eq.~\eqref{eq:methods:homodyne_difference_current} can be rewritten as:
\begin{equation}
I^-_\mathrm{HD} = K \eta E_\mathrm{LO} (E_s^*e^{i\phi} + E_s e^{-i\phi}).	
\label{eq:methods:homodyne_difference_current2}
\end{equation}
Readily, at $\phi = 0$ and $\phi = \pi/2$ the Faraday rotation and the ellipticity noise are measured, respectively, as seen from the direct
comparison of Eqs.~\eqref{eq:methods:faraday_difference_current},
\eqref{eq:methods:ellipticity_difference_current}, and \eqref{eq:methods:homodyne_difference_current2}.


\begin{thebibliography}{}
\bibitem{CommentBHD} NewFocus 1607 balanced receiver with two Si photodiodes provided the noise-equivalent power of 40~pW/$\surd{}$Hz and gain of 350~V/W is used.
\end{thebibliography}
\end{document}